# Exploring the bouncing cosmological models in symmetric teleparallel gravity

**A.Y.Shaikh***

*Department of Mathematics, Indira Gandhi Mahavidyalaya, Ralegaon 445402, India.

**e-mail:** shaikh_2324ay@yahoo.com

**Abstract**

In this study, cosmological models with perfect fluid and a gravitational framework $f(Q)$ will be examined. In this modified theory of gravity, the gravitational force has the form $f(Q)$, where $Q$ stands for the non-metricity scalar. I create two bouncing cosmological models, one in which the Lagrangian $f(Q)$ is assumed to have a linear dependence on $Q$ and the other in which it has a polynomial functional form. It has been discovered that the parameters of the individual models largely determine how they will behave. The resulting models' equation of state (EoS) parameter captures the universe's ironic behavior. It should be highlighted that the built-in cosmological models go against the energy requirements. The models' kinematical and physical characteristics are discussed.

# I. Introduction

Numerous cosmological discoveries, including supernova, CMB radiation anisotropy, large-scale structure, and baryon acoustic oscillation, have demonstrated that the universe is expanding more quickly, at least in its advanced stages of evolution. Modified gravity theories and dark energy models are two distinct explanations for the cause of the Universe's current acceleration. Dark energy (DE) is analogous to an enigmatic fluid with negative pressure that defies gravity and speeds up the Universe's expansion. The cosmological constant that characterizes the $\Lambda$ dominant Cold Dark Matter paradigm is the most straightforward DE candidate ($\Lambda$ CDM). In the literature, a number of modified ideas have been put forth, including theories $f(R)$ [1-3], $f(T)$ [4-6], $f(T,B)$ [7], $f(R,T)$ [8-9], $f(Q,T)$ [10-11], $f(G)$ [12], and $f(R,G)$ theory concept [13–14], etc. Gravitational ideas have been extensively studied in the modern era. The concept of symmetric teleparallel gravity was first presented by J.B. Jiménez et al. [15]. The $f(Q)$ theory also functions as a teleparallel gravity-like alternative to general relativity. The non-metricity $Q$ describes gravitational interactions in symmetric teleparallel gravity. T. Harko investigated the symmetric teleparallel gravity's expansion in Ref. [16]. Noemi conducted an inspiring investigation into $f(Q)$ gravity in which he looked at the fundamental non-metricity gravity signals [17]. A strong set of constraints on $f(Q)$ gravity are imposed by Lazkoz et al. [18], where Lagrange is given as a polynomial function of redshift z. When subjected to the restrictions of the energy conditions as indicated in [19], the $f(Q)$ model likewise revealed a comparable description of an accelerated phase. Khyllep et al. [20] have demonstrated the cosmological viability of the $f(Q)$ gravity model by performing the singularity analysis and dynamical system analysis. According to Barros et al. [21], the tension between Planck and LSS data can be reduced within this framework by analyzing the linear matter fluctuations are numerically evolved and the examination of the growth rate of structures. The first indication that the non-metricity of $f(Q)$ gravity might conflict with the Universe's LCDM behavior comes from Anagnostopoulos et al. [22]. In order to understand the Universe's accelerated expansion within the context of changed gravity, R. Solanki et al. [23] have looked into the function of bulk viscosity. Two accelerating cosmic models with symmetric teleparallel $f(Q)$ gravity were introduced by Narwade et al. [24]. For a $f(Q)$ polynomial model, Dimakis et al. researched quantum cosmology [25]. Numerous contexts and cosmological models of $f(Q)$ gravity have been thoroughly investigated (see references there in [26-31]).

In the current project, I investigated a few bouncing models inside the symmetric teleparallel gravity theory framework. Our objective is to explain the late-time cosmic speed-up event and analyze the bouncing behavior at an initial epoch using a straightforward symmetric teleparallel gravity theory that explores geometrical degrees of freedom.

## II. Motion Equations in $f(Q)$ gravity

Take into account the action for $f(Q)$ provided by

$$S = \int \frac{1}{2} f(Q) \sqrt{-g} \, d^4x + \int L_m \sqrt{-g} \, d^4x, \qquad (1)$$

where $f(Q)$ is a generic function of the $Q$, $L_m$ is the matter Lagrangian density and $g$ is the metric determinant. The traces of the nonmetricity tensor are such that

$$Q_{\gamma\mu\nu} = \nabla_\gamma g_{\mu\nu}, \qquad (2)$$

$$Q_\gamma = Q_\gamma{}^\mu{}_\mu, \quad \tilde{Q}_\gamma = Q^\mu{}_{\gamma\mu}. \qquad (3)$$

The nonmetricity tensor trace has the following form $Q = -Q_{\gamma\mu\nu} P^{\gamma\mu\nu}$. (4)

The energy momentum tensor for the matter, whose definition is, is another important component of our strategy. Finding the field equations 
$$T_{\mu\nu} = -\frac{2}{\sqrt{-g}} \frac{\delta(\sqrt{-g} L_m)}{\delta g^{\mu\nu}} \tag{5}$$

requires taking the action (1) variation with regard to the metric tensor,

$$\frac{2}{\sqrt{-g}} \nabla_\gamma \left( \sqrt{-g} f_Q P^\gamma{}_{\mu\nu} \right) + \frac{1}{2} g_{\mu\nu} f + f_Q \left( P_{\mu\dot\gamma} Q_\nu{}^{\dot\gamma} - 2 Q_{\dot\gamma\mu} P^{\dot\gamma}{}_\nu \right) = T_{\mu\nu}, \tag{6}$$

where $f_Q = \frac{df}{dQ}$. In addition, we may also consider (1)'s modification in terms of the relationship, leading to $\nabla_\mu \nabla_\gamma \left( \sqrt{-g} f_Q P^\gamma{}_{\mu\nu} \right) = 0$. \tag{7}

A flat FRW metric has the form $ds^2 = -dt^2 + a^2(t)(dx^2 + dy^2 + dz^2),$ \tag{8}

where $a(t)$ is a function of $t$ and is the scale factor. We may express the trace of the nonmetricity tensor as $Q = 6\left(\frac{\dot a}{a}\right)^2$ recognitions to the line element. Take a look at the energy-momentum tensor for a perfect fluid, or $T_{\mu\nu} = (\rho + p) u_\mu u_\nu + p g_{\mu\nu},$ \tag{9} where $p$ and $\rho$ are the pressure and energy density, respectively. Therefore, one can find by substituting (9), and (8) in (6)

$$3\left(\frac{\dot a}{a}\right)^2 = \frac{1}{2 f_Q}\left(-\rho + \frac{f}{2}\right), \tag{10}$$

$$\frac{\ddot a}{a} + 2\left(\frac{\dot a}{a}\right)^2 + \frac{\dot f_Q}{f_Q}\left(\frac{\dot a}{a}\right) = \frac{1}{2 f_Q}\left(p + \frac{1}{2}\right), \tag{11}$$

as the gravity-specific modified Friedmann equations. Here, (.) stands for one time-related derivative.

## III. Cosmological Model of Bouncing

According to [32–39], the Universe has an initial reduction phase in which matter is subdued, followed by a non-singular bounce. According to Ref. [40], a bouncing Universe is one in which the universe first collapses, then reaches a minimum, and finally expands. In bouncing cosmology, the scale factor increases $(\dot{a}>0)$ during the expanding phase (positive time zone) and decreases $(\dot{a}<0)$ during the contracting phase (negative time zone). It yields $(\dot{a}=0)$ at the bouncing point. Let's think about a symmetric bounce through the scale factor $a = e^{\gamma t^2}$, (12) where $\gamma$ is a positive constant parameter that governs the growth of the Universe. The Hubble parameter can be found by using $H = \dfrac{\dot{a}}{a} = 2\gamma t.$ (13)

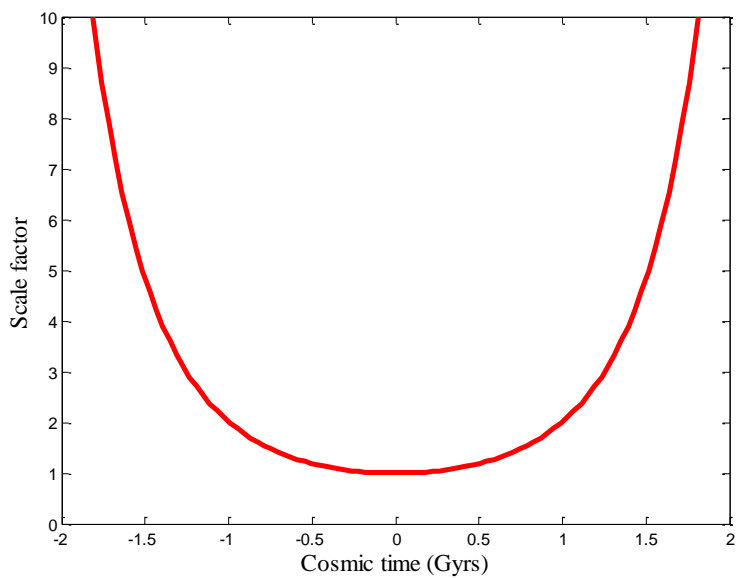

Figure 1 shows a plot of the scale factor's variation across cosmic time.

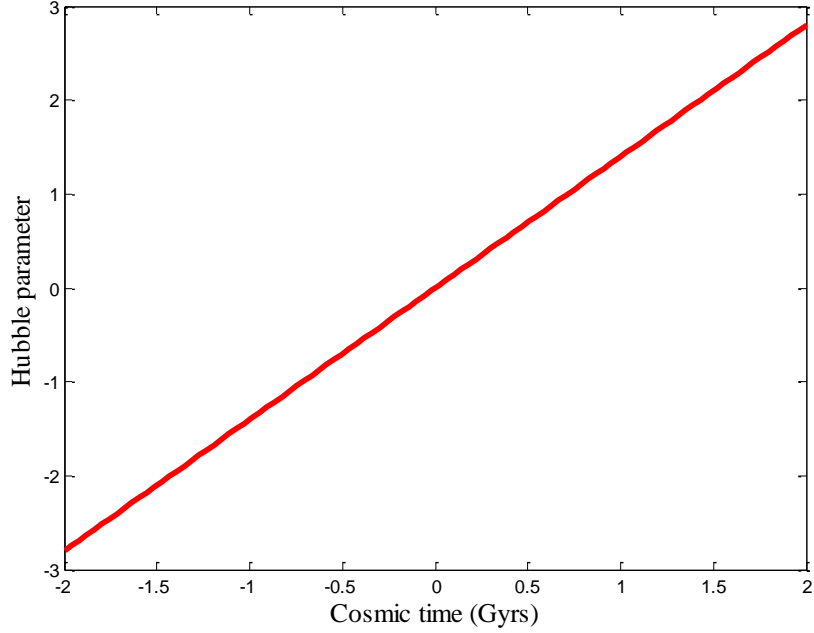

Figure 2. A plot showing the Hubble parameter's fluctuation across cosmic time.

The scale factor's fluctuation with regard to cosmic time is depicted in Fig. 1. The scale factor behaves symmetrically at the point of bounce $t=0$. The parameter's value $\gamma$ expresses the angle at which the scale factor's curves are curved. Since the scale factor value is continuous (does not disappear) at the bouncing point $t=0$, it is possible for the Hubble parameter to have a null value during the bouncing epoch. To fully comprehend a bounce, consider the contrast between an expanding Universe $H>0$ and one that depicts a contracting cosmos $H<0$. Consequently $\dot{H}>0$, this is seen at and around the bounce point. The Hubble parameter's linear fluctuation with respect to cosmic time is seen in Figure 2. The Hubble parameter advances linearly from the contracting Universe in relation to the bouncing model. For this bouncing scale factor, the deceleration parameter is defined and may be derived as $q=-1-\dfrac{\dot{H}}{H^2}=-1-\dfrac{1}{2\gamma t^2}.$ (14)

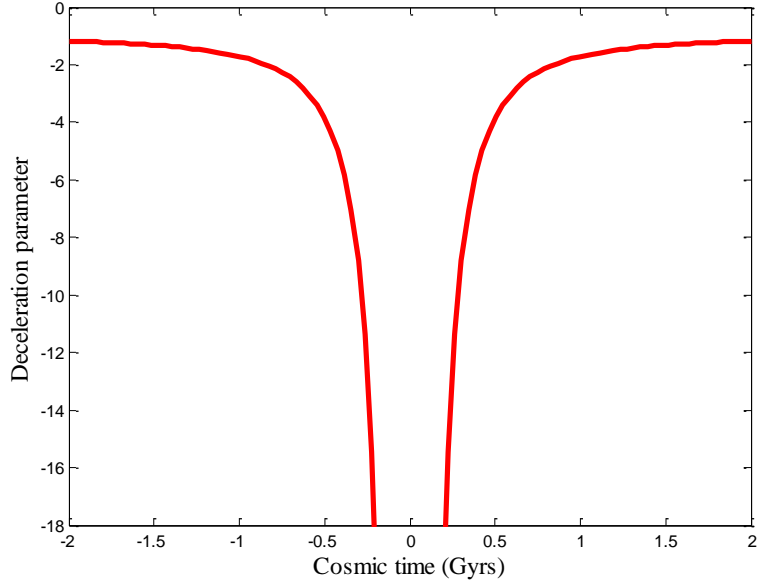

Figure 3 shows a plot of the deceleration parameter fluctuation throughout cosmic time. The deceleration parameter sheds light on how the model expands. According to Ref. [41], cosmological models are divided into the following categories based on their time dependence on the Hubble parameter and deceleration parameter: i) $H>0, q>0$; expanding and decelerating; (ii) $H>0, q<0$; expanding and accelerating; (iii) $H<0, q>0$; contracting and decelerating; (iv) $H<0, q<0$; contracting and accelerating; (v) $H>0, q=0$; expanding, zero deceleration/constant expansion; (vi) $H<0, q=0$; contracting, zero deceleration; and (vii) $H=0, q=0$; static. The symmetrical behavior of the deceleration parameter at the bouncing point $t=0$ is shown in Fig. 3. It is important to note that the deceleration parameter has a negative value for both expanding and contracting universes, and that over a limited amount of time, it tends to have a constant value of -1. In the negative time zone (contacting Universe), even after developing from, the deceleration parameter tends to significant negative values near the bouncing point. The deceleration parameter tends to $q=-1$ at late times in the expanding Universe's positive cosmic time frame. According to the values of the deceleration parameter, the authors of [41] explored the different types of expansion that our Universe exhibits, including i) $q < -1$; super exponential expansion, (ii) $-1 \leq q < 0$; exponential expansion ($q = -1$ also known as de-Sitter expansion), (iii) $q = 0$; expansion with constant rate, (iv) $-1 < q < 1$; accelerating power expansion, and (v) $q > 0$; decelerating expansion. The cosmos is therefore expanding faster than previously thought and the deceleration parameter value agrees with recent cosmological findings of Type Ia Supernovae.

### III(a): Model I

Let's have a look at a linear Lagrangian $f(Q)$ i.e. $f(Q) = \alpha Q$, where $\alpha$ is just a random constant.

The energy density is given by $\rho = -12\alpha\gamma^2 t^2$. (15)

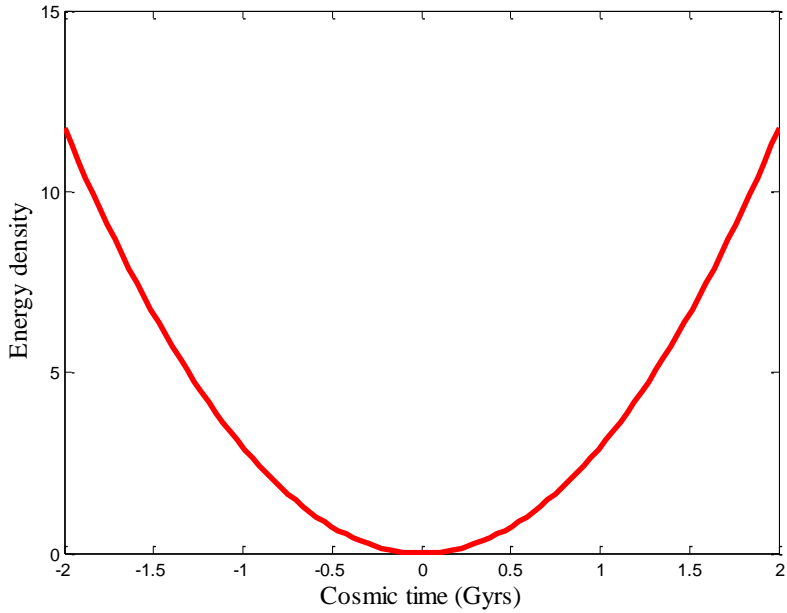

Figure 4. Plot showing the change in energy density across cosmic time.

This is how the pressure is measured $p = 36\alpha\gamma^2 t^2 - 4\alpha\gamma$. (16)

The parameter of the Equation of State (EoS) is written as $\omega = \frac{p}{\rho} = -3 + \frac{1}{3\gamma t^2}$. (17)

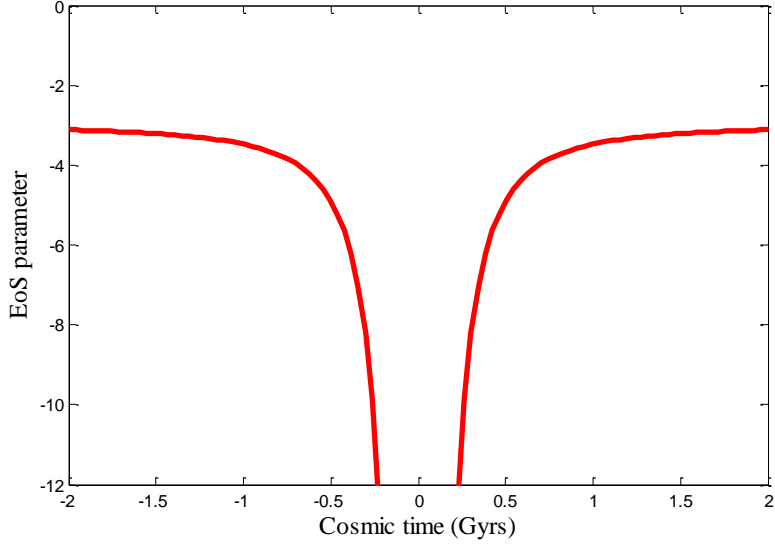

Figure 5 shows a plot of the EoS parameter variation across cosmic time.

We are aware that energy density $\rho$ should increase as the universe expands. Therefore, all the parameters used to explain the resulting model may be constrained using this criterion $\rho \geq 0$. As a result, it is seen that for the first model, the energy density will be positive for $\alpha < 0$ throughout the course of the Universe's evolution. The equation of state parameter, it should be noted, $\omega \leq -1$ indicates the phantom phase close to the bouncing point.

### III (b) :- Model II

Let's have a look at a non-linear Lagrangian $f(Q)$ i.e. $f(Q) = Q + mQ^n$, where $m$ and $n$ are just some random constants.

The energy density is given by $\rho = -12\alpha\gamma^2 t^2 + \left(\frac{1}{2}m - mn\right)\left(24\gamma^2 t^2\right)^n$. (18)

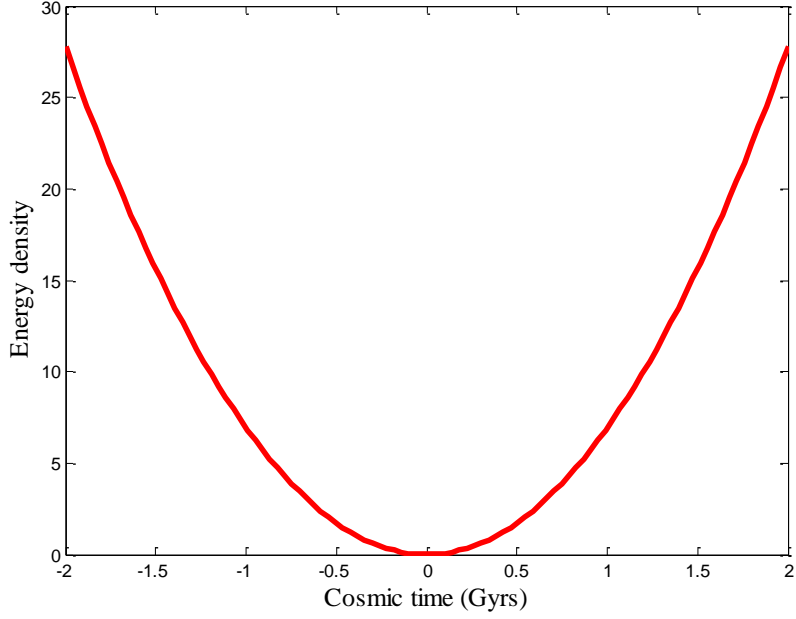

Fig.6. Plot showing how energy density varies with cosmic time.

The pressure is written as

$$p = 36\alpha\gamma^2 t^2 + \left(\frac{1}{2}m - mn\right)\left(24\gamma^2 t^2\right)^n - 4\gamma \\ - 4\gamma mn\left(24\gamma^2 t^2\right)^{n-1} + 8\gamma mn(n-1)\left(24\gamma^2\right)^{n-1} t^{2n-2}. \tag{19}$$

The Equation of State ( EoS ) parameter is expressed as

$$\omega = \frac{36\alpha\gamma^2 t^2 + \left(\frac{1}{2}m - mn\right)\left(24\gamma^2 t^2\right)^n - 4\gamma - 4\gamma mn\left(24\gamma^2 t^2\right)^{n-1} + 8\gamma mn(n-1)\left(24\gamma^2\right)^{n-1} t^{2n-2}}{-12\alpha\gamma^2 t^2 + \left(\frac{1}{2}m - mn\right)\left(24\gamma^2 t^2\right)^n}. \tag{20}$$

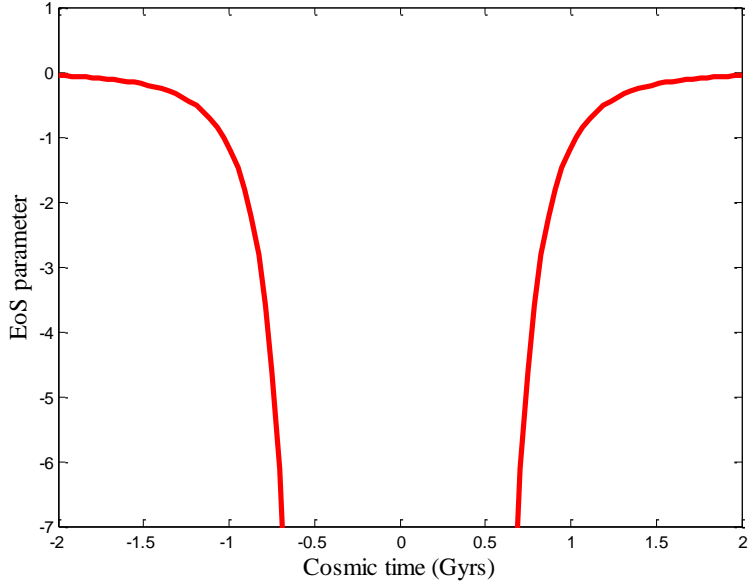

Fig.7. A plot showing how the EoS parameter changes throughout cosmic time.

We are aware that as the Universe expands, energy density $\rho$ should increase. Therefore, all the parameters used to explain the resulting model may be constrained using this criterion $\rho \geq 0$. As a result, for the second model, it is found that for $n \geq 0.8$ and $m \leq 0$ during the course of the Universe's evolution, the energy density will be positive. The equation of state parameter, it should be noted, $\omega \leq -1$ indicates the phantom phase close to the bouncing point.

## IV. Statefinder Diagnostic

Sahini et al. and Alam et al. built the state finder diagnostic pair in Refs. [42-43] using the second and third derivatives of the scale factor, where $r$ denotes the jerk parameter and $s$ denotes the material of the DE i.e. $r = \dfrac{\dddot{a}}{aH^3}$, $s = \dfrac{(r-1)}{3\left(q - \dfrac{1}{2}\right)}$. (21)

The model exhibits cold dark matter (CDM) limit behavior $(r,s) = (1,1)$ while $(r,s) = (1,0)$ producing $\Lambda$CDM limit contributions. When $r<1$, it produces a quintessence DE region, whereas for $s>0$ phantom DE areas. Statefinder parameters $r = 1 + \frac{3}{2\gamma t^2}$ and $s = -\frac{1}{1+3\gamma t^2}$ are a pair. Given by

$$r = 1 - \frac{9\gamma s}{2\gamma(1+s)}, \qquad (22)$$

the relationship between $r$ and $s$.

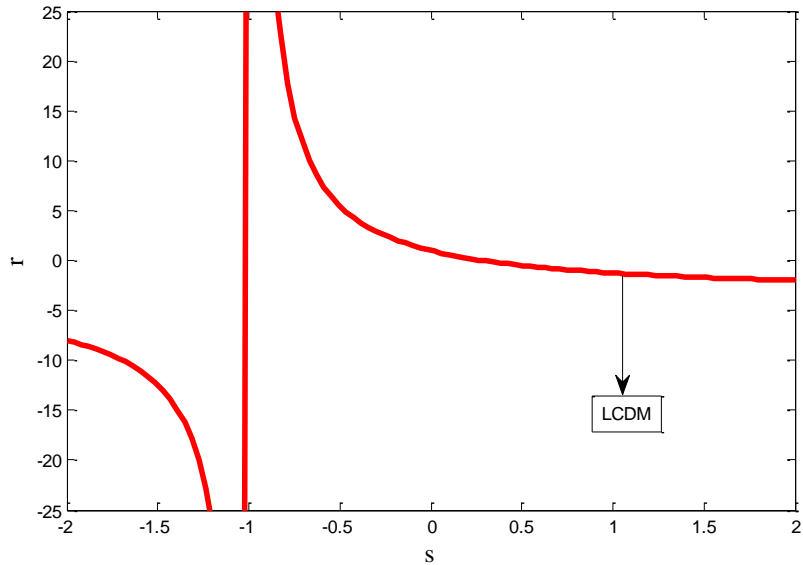

Fig.8. Plot illustrating the fluctuation of *r* against *s*.

It is important to notice that, according to equation (22), the model acts like a flat model over infinite time, which will ultimately result in $s = 0$ and $r = 1$. Thus, it has been demonstrated that the $\Lambda$CDM flat model will cause the model to occur instantly. Figure 8 illustrates the statefinder pair's behavior. It appears as though $s > 0$ and $r < 1$ belong in this area. The derived model thereafter agrees with the DE regions like quintessence and phantom.

## V. Jerk parameter

The third derivative of the scale factor pertaining to cosmic time is how the jerk parameter is defined $j(t) = \frac{1}{H^3}\frac{\ddot{a}}{a} = q + 2q^2 - \frac{\dot{q}}{H}$. The jerk parameter in cosmology is used to confer the models close to the $\Lambda$CDM. The phase conversion from deceleration to acceleration is caused by the negative value of the deceleration and the positive value of the jerk parameter. The value $j = 1$ is present for the $\Lambda$CDM model's jerk parameter. The value of the jerk parameter is

$$j(t) = 1 + \frac{3}{2\gamma t^2}. \tag{23}$$

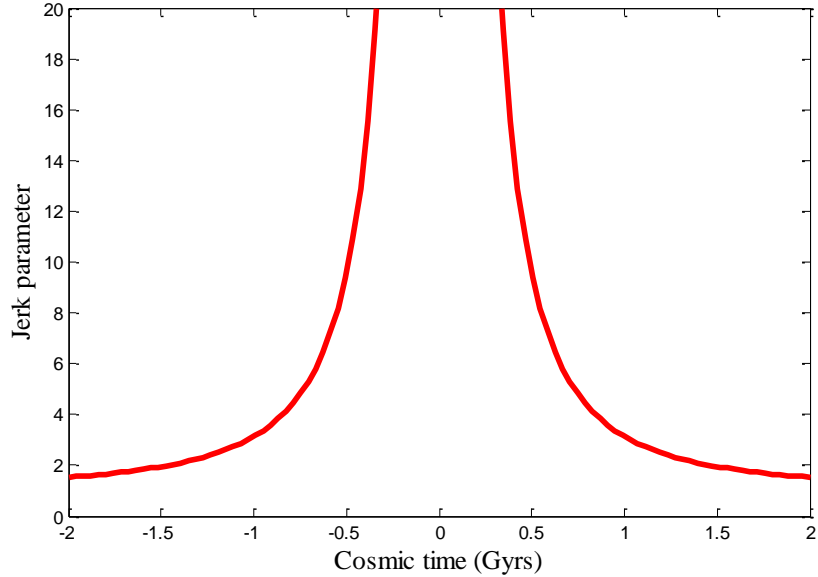

Figure 9. Plot showing the jerk parameter's fluctuation across cosmic time.

Figure 9 depicts the jerk parameter graphically in relation to cosmic time. The value of the parameter $\gamma$ affects the jerk parameter. The jerk parameter has a solitary value as a result of the Hubble rate's disappearing nature at the bouncing point. The jerk parameter decreases to negligibly small values around one, endlessly far from the bounce.

## VI. Energy Situations

Energy conditions (ECs) are a set of linear equations combining pressure and density that show that gravity is always repulsive and that energy density cannot be negative. These four types of ECs are expressed as follows:

(1) Null Energy condition (NEC) $\Leftrightarrow \rho + p \geq 0$ (24)

(2) Strong Energy Condition (SEC) $\Leftrightarrow \rho + 3p \geq 0$ (25)

(3) Dominant Energy Condition (DEC) $\Leftrightarrow \rho \geq |p|$ (26)

(4) Weak Energy Condition (WEC) $\Leftrightarrow \rho \geq 0, \rho + p \geq 0$ (27)

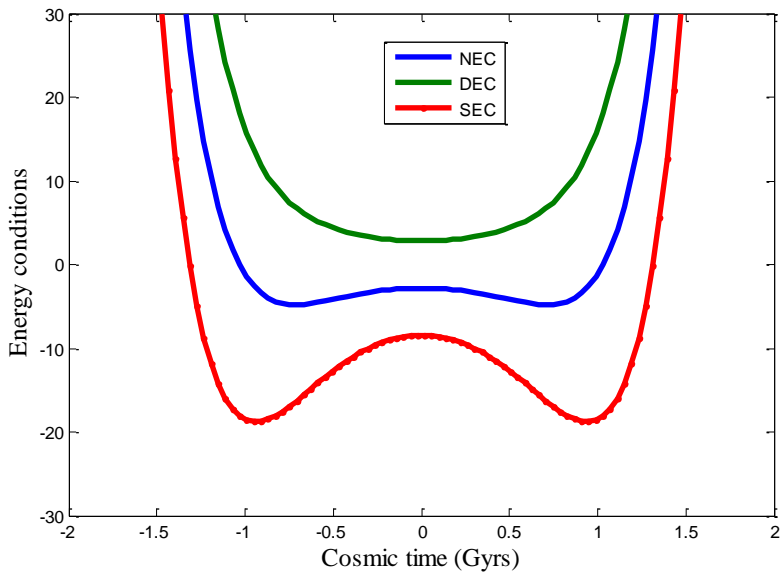

Fig.10. Plot for the variation of the energy conditions versus cosmic time for Model I.

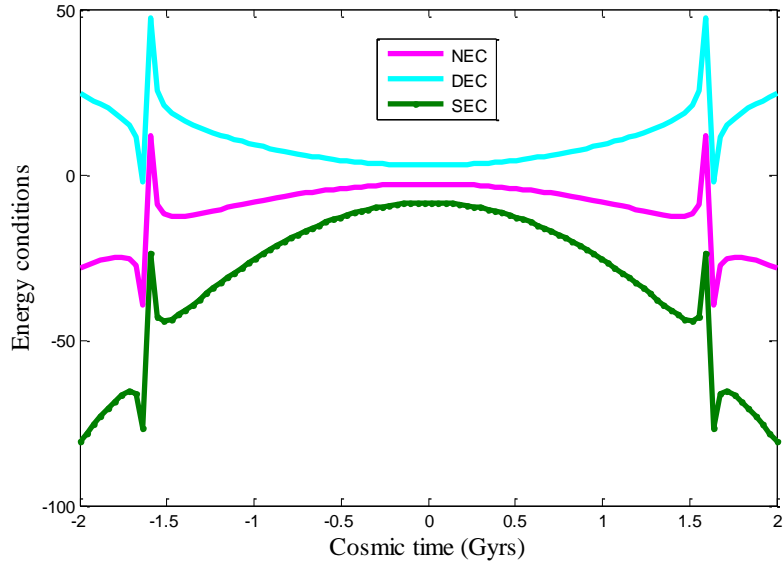

Fig.11. Plot for the variation of the energy conditions versus cosmic time for Model II.

Figures 10 (Model I) and 11 (Model II) depict the NEC, DEC and SEC against cosmic time for the current bouncing model, and as predicted, both the NEC and SEC are broken at the bouncing point. In actuality, the scenario of bouncing involves a basic breach of the energy condition. Figures 10 and 11 show that the SEC violation that is causing the emergence of dark energy. The early description's singularity problem is resolved by the observation that the NEC and SEC are non-singular at the bouncing point, which also serves to illustrate a non-singular bouncing Universe. The energy conditions are obtained as $\rho + p \leq 0$ and $\rho + 3p \leq 0$ at the bouncing point, thus $\rho + p \geq 0$ and $\rho + 3p \geq 0$ violate the energy criteria. The SEC condition must be broken in order to describe a Universe that is governed by negative pressure [44]. According to the most recent data on the expanding Universe, the SEC must be broken on an astronomical scale [45–48]. The bad behavior of SEC shows how quickly the Universe is expanding.

# VII. Discussion and final thoughts

According to the bouncing Universe cosmology, the cosmos begins to contract and twitches to expand in order to avoid any singularities (see References [49–51]). In this study, I took into account the exact shape of the Hubble parameter, which is a function of cosmic time $t$, in the flat FLRW model. From a contracting to an expanding cosmos, the Hubble parameter moves linearly through a bounce. The deceleration parameter advances to significant negative values close to the bounce in the negative domain. It shifts from extremely negative values to late times in the positive domain. $\rho + p \leq 0$ and $\rho + 3p \leq 0$ exhibit a breach of the strong energy condition and null energy condition, demonstrating the Universe's late-time acceleration. Furthermore, it has been discovered that the cosmic jerk parameter has been positive throughout the whole history of the Universe and that it approaches to 1 in later periods. The main advantage of the bouncing cosmology is that it provides a method for addressing the singularity problem that appears in the conventional astrophysical model.